\documentclass{aa} 
\usepackage{graphicx}
\usepackage{natbib}
\bibpunct{(}{)}{;}{a}{}{,}
\usepackage{txfonts}
\begin{document}
   \title{The peculiar galaxy Mkn 298 revisited through integral field
        spectroscopy
\thanks{Based on data obtained at the  2.2m and NTT telescopes of ESO-la 
Silla (Chile), at the  2.2m telescope of DSAZ-Calar Alto (Spain), and at the 
6m telescope of SAO (Russia).}
}

   \author{M. Radovich\inst{1}
      	\and
        S. Ciroi\inst{2}
      	\and
        M. Contini\inst{3,2}
	\and      
        P. Rafanelli\inst{2}
      	\and 
        V. L. Afanasiev\inst{4,1}
      	\and
        S. N. Dodonov\inst{4,1}
        }

   \institute{
	INAF, Osservatorio Astronomico di Capodimonte, Via Moiariello 16, 
	I-80131, Napoli, Italy\\ 
		\email{radovich@na.astro.it}
	\and	
	Dipartimento di Astronomia, Universit\`a di Padova, 
	Vicolo dell'Osservatorio 2, I-35122  Padova, Italy\\	 
        	\email{ciroi@pd.astro.it, piraf@pd.astro.it}
	\and
	School of Physics and Astronomy, Tel-Aviv University,
             	Ramat-Aviv, Tel-Aviv, 69978 Israel\\
             	\email{contini@post.tau.ac.il}
	\and
	Special Astrophysical Observatory, Nizhnij Arkhyz, 369167 Russia\\
		\email{vafan@sao.ru, dodo@sao.ru}	     
        }

   \offprints{M. Radovich}


\date{Received; accepted}

\abstract{ Spectroscopic and imaging data of the peculiar galaxy Mkn 298 
are presented in this paper. Narrow-band \ion{H}{$\alpha$} and broad-band
$R$ images are used to study  the star formation rate in the galaxy
and its morphology, which is typical of a merging system.
Long-slit and integral field spectra are used to assess the
kinematics of gas and stars, and the 
nature of the ionizing source at different distances from the nucleus. 
In particular, the nucleus of Mkn 298 is characterized by peculiar line 
ratios: 
\ion{[N}{ii]}$\lambda6583$/\ion{H}{$\alpha$} is typical of \ion{H}{ii}-like
regions, while \ion{[O}{i]}$\lambda6300$/\ion{H}{$\alpha$} could indicate the 
presence of an active galactic nucleus. 
We show that models where a shock component is  added to  photoionization 
from a starburst allow to reproduce the observed line ratios.  
Mkn 298 is thus most likely a star forming galaxy, rather than a galaxy
 hosting an active nucleus.  
}

\titlerunning{The peculiar galaxy Mkn 298}
\authorrunning{M. Radovich et al.}   
    
   \maketitle

%

\def\ea{\it et al. \rm}
\def\am{$^{\prime}$\ }
\def\as{$^{\prime\prime}$\ }
\def\msol{M$_{\sun}$ }
\def\kms{$\rm km\, s^{-1}$}
\def\cm3{$\rm cm^{-3}$}
\def\Ts{$\rm T_{*}$}
\def\Vs{$\rm V_{s}$}
\def\n0{$\rm n_{0}$}
\def\B0{$\rm B_{0}$}
\def\ne{$\rm n_{e}$}
\def\Te{$\rm T_{e}$}
\def\Tgr{$\rm T_{gr}$}
\def\Tgas{$\rm T_{gas}$}
\def\Ec{$\rm E_{c}$}
\def\Fh{$\rm F_{H}$}
\def\Hb{\ion{H}{$\beta$}}
\def\Ha{\ion{H}{$\alpha$}}
\def\Fn{$\rm F_{n}$}
\def\Fh{$\rm F_{h}$}
\def\erg{$\rm erg\, cm^{-2}\, s^{-1}$}
\def\mum{$\mu$m}
\def\Lx{L$_X$~}
\def\Fx{F$_X$}
\def\LIR{L$_{IR}$~}
\def\L12{L$_{12\mu m}$~}
\def\F12{F$_{12\mu m}$~}
\def\agr{a$_{gr}$}
\def\mm{$\mu$m}

---------------------------------------------------------------------

\section{Introduction}

\object{Mkn 298} (IC 1182) is known as a morphologically peculiar system
(Figure~\ref{fig:bim}) located in the Hercules supercluster.  
It shows a chain (c, d, e) of small compact blue 
emitting regions \citep{stock} aligned on its eastern side up
to $\sim 80$\arcsec\ from the main body of the galaxy, which is classified
in NED (NASA/IPAC Extragalactic Database) as a SA0+ galaxy.
These regions are aligned with a tail (b) resembling
a spiral arm located on the eastern side of the galaxy (a). 
On the western side a very faint trace of spiral arm is also visible.
A faint emission is visible in the c knot, behind a foreground star 
\citep{stock,met:pron}. 
The d and e knots are  characterized by an emission line 
spectrum. They were identified by \citet{braine} as a tidal 
dwarf galaxy (TDG hereafter) forming from 
material ejected from the disk of a galaxy after a collision.

The classification of Mkn 298 as a galaxy hosting an 
active nucleus is controversial in literature.
\citet{koski} and \citet{vo} classified it as a Seyfert 2
galaxy with an extremely low \ion{[N}{ii]}$\lambda$6583/\ion{H}{$\alpha$} line 
ratio. 
Conversely \citet{vieg:gruen} enclosed Mkn 298 in a sample of LINERs.
Variability was claimed by \citet{met:pron} who
observed  changes of the relative intensity of 
\ion{[O}{iii]}$\lambda\lambda$4959,5007 and \ion{H}{$\beta$} in the period
1969--70, 1977. The detection  of variable 
soft X-ray emission \citep{rafanelli} with a luminosity typical of
Seyfert 2 galaxies,  seems to confirm the presence of an active nucleus.
However, as noted by \citet{moles} it is not possible to exclude that the 
X-ray emission is produced in star-forming regions \citep{zezas}.
Finally, \citet{moles} detected the presence of several knots in the 
nucleus; they interpreted Mkn 298 as a very luminous starburst (SFR $\sim$
90 $M_{\sun}$/yr) with very low metallicity ($Z/Z_{\sun} \sim 0.06 - 0.1$).

An extended \ion{H}{i} distribution peaked on the TDG was detected  using VLA 
data \citep{braine,iglesias}. 
\citet{braine} report the detection of CO(2-1) and CO(1-0) emission in Mkn 298, 
however the emission is peaked on the main galaxy and is below detection in 
the  TDG. They computed a total amount of H$_2$ 
mass in the system $M_{\rm mol} > 5 \times 10^9$ $M_{\sun}$, and an upper 
limit $M_{\rm mol} <6 \times 10^7$ $M_{\sun}$ in the TDG. 
This would imply that  only a small fraction of HI was transformed 
into H$_2$ in the TDG, while in the nucleus the conversion already took place.
This is consistent with Mkn 298 being in the late stage of a merging: as 
outlined by \citet{hibbard}, in advanced mergers the
atomic gas is mostly relegated to the outer tidal features and is almost 
absent in the central regions, whereas the molecular gas is concentrated in 
the remnant body.

The aim of this paper is to study in detail the physical conditions
in the different components (galaxy body and tail) of the Mkn 298 system
by the analysis of their emission-line spectra. 
Observations and data reduction steps are described in 
Sec.~\ref{sec:observations}.
The morphology, star formation rates and kinematics are analyzed and 
discussed in  Sec.~\ref{sec:morphology},  Sec.~\ref{sec:sfr} and  
Sec.~\ref{sec:kinematics} respectively.   
In Section~\ref{sec:linerat} emission line ratios are used to derive
chemical abundances. Observed ratios are then compared with the results 
from both pure  photoionization models and photoionization+shock models,
using the {\sc Cloudy} and {\sc Suma} codes respectively. 
Conclusions are finally drawn in Sec.~\ref{sec:conclusions}.

\section{Observations and data reduction\label{sec:observations}}
Emission line spectra and images of Mkn 298 have been taken in different 
periods and using different observing facilities.

\subsection{Imaging}

\begin{enumerate}
\item
Narrow band images in \ion{H}{$\alpha$} ($\lambda_c$=6820 \AA, $\Delta \lambda$=170 \AA) and 
continuum light ($\lambda_c$=6240 \AA, $\Delta \lambda$=110 \AA) were taken at 
the Deutsch-Spanische Astronomische Zentrum of Calar Alto (DSAZ) 
2.2~m telescope (Spain) in September 1996, with a 2k$\times$2k 
CCD (spatial scale 0.53\arcsec/px), and under a seeing of $\sim 1.5$\arcsec.
Images of the galaxy of 600 and 900 s were obtained with the 
two filters, respectively. Moreover, the spectrophotometric standard star 
LDS749b was observed for flux calibration.
All images were processed by means of the IRAF\footnote{IRAF is distributed 
by the National Optical Astronomy Observatories, which are operated by the 
Association of Universities for Research in Astronomy, Inc., under cooperative 
agreement with the National Science Foundation} packages. They were firstly 
bias subtracted, flat-fielded and cosmic-rays cleaned.
Then, we performed the photometry of the standard star, whose fluxes 
were measured with PHOT, corrected for atmospheric extinction, and compared to
the fluxes obtained from the spectrum of the star convolved with the response
curve of the filters, in order to obtain the calibration constants. 
The filter transmissions were conveniently taken into account.
The images of the galaxy were aligned, background subtracted, normalized to 1 
s exposure and corrected for atmospheric extinction. Then, the continuum 
image was scaled by a factor 1.6 to compensate the different shapes of the two 
filters. The respective calibration constants were applied to the so processed 
images, and finally the continuum was subtracted from \ion{H}{$\alpha$} 
to obtain a pure emission line image.

\item
Five broad-band R consecutive exposures of Mkn 298, 15 min each, 
were obtained with EMMI at the ESO-NTT in August 1991. 
The 2k$\times$2k CCD, with 15$\mu$-size pixels, yielded a spatial scale of 
0.348\arcsec/px. The seeing was 1.4\arcsec. 
The five images were bias and flat-field corrected, then aligned and 
combined to obtain an average frame cleaned from cosmic rays.
No photometric calibration was performed.

\item
A public image of the galaxy was extracted from the HST Archive.
Obtained in June 1995 with the WFPC2 in combination with the F606W filter, and 
500 sec of exposure time \citep[see][]{malkan}, this image 
was simply cleaned from  cosmic-rays and directly analyzed. 
The main part of the galaxy was covered with the PC1, having a spatial scale 
of 0.045\arcsec/px.

\end{enumerate}

\subsection{Spectroscopy}

\begin{enumerate}

\item The nucleus  was observed in May and June 1999 with the Multi Pupil Fiber
Spectrograph (MPFS) at the 6~m telescope of the Special Astrophysical
Observatory (SAO RAS, Russia).
The MPFS is an integral field spectrograph, in which an array
(16$\times$15 at the epoch of the observations)
 of 1\arcsec$\times$1\arcsec microlenses  is connected to a bundle of fiber 
optics.
The 240 fibers are re-arranged to form a pseudo-slit which is the input to 
the spectrograph. 
A 600/mm grating was used during the first run (R1) of observations,
in combination with a 1k$\times$1k CCD.
Each of the two  spectral ranges, 3350-6100 \AA\ and 4550-7300 \AA, 
was observed  with a dispersion of $\sim 2.6$ \AA/px,  an instrumental 
FWHM of 7.5 \AA\  ($\sim 400$ \kms at 5500 \AA) and an exposure time of 3600 
sec.
During the second run (R2) a 1200/mm grating was used, which covered the 
ranges 4800-6200 \AA\ and 5900-7300 \AA, with a dispersion of $\sim 1.36$ 
\AA/px and an 
instrumental FWHM of 4.5 \AA\  ($\sim 250$ \kms at 5500 \AA).
Integration times of 7200 sec for the ``blue spectrum'' and 4800 sec for the 
``red spectrum'' were applied. 
The seeing was typically around 1.5-2\arcsec\ for both runs. 

The MPFS spectra were reduced and flux calibrated with a special software 
developed in IDL at SAO. 

\item  Knots d,e were also observed with the MPFS during the above
runs. In this case, a 300/mm grating was used, which allowed 
to cover
the whole optical spectral range 3600-9000 \AA, with an instrumental FWHM 
$\sim 15$ \AA. An exposure time of 3600 sec was applied.
Long-slit spectra of the same knots were taken with EFOSC2 at the 2.2-m
telescope at ESO-La Silla (Chile) in April 1995.
The instrumental FWHM is $\sim 10$ \AA, the covered range is 
$4500-7000$ \AA. 
The standard data reduction steps (bias subtraction, flat-field correction
and flux calibration) were applied using the LONGSLIT package in IRAF.
Due to the higher spectral resolution of the ESO spectra, we decided to
use them rather than the MPFS data in the analysis of the knots, with the
exception of the \ion{[O}{ii]}$\lambda3727$ line which is covered by 
the MPFS spectra only.

\end{enumerate}

\begin{figure}
\includegraphics[width=60mm,angle=270]{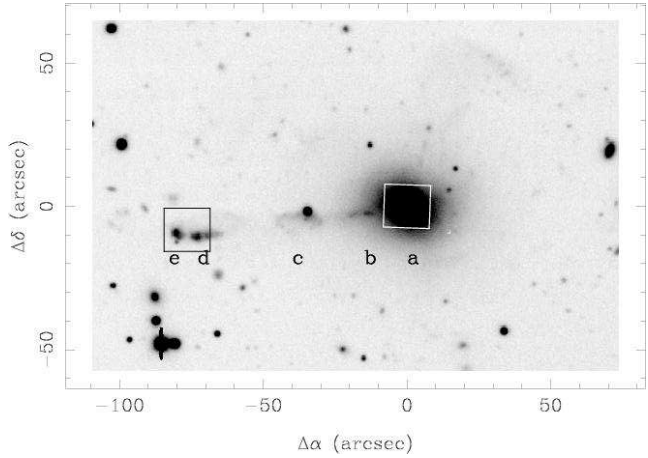}
\caption
  {The Mkn 298 complex  in the $R-$band image with overlaid the 
position of the MPFS arrays.\label{fig:bim} }
\end{figure}

\begin{figure}
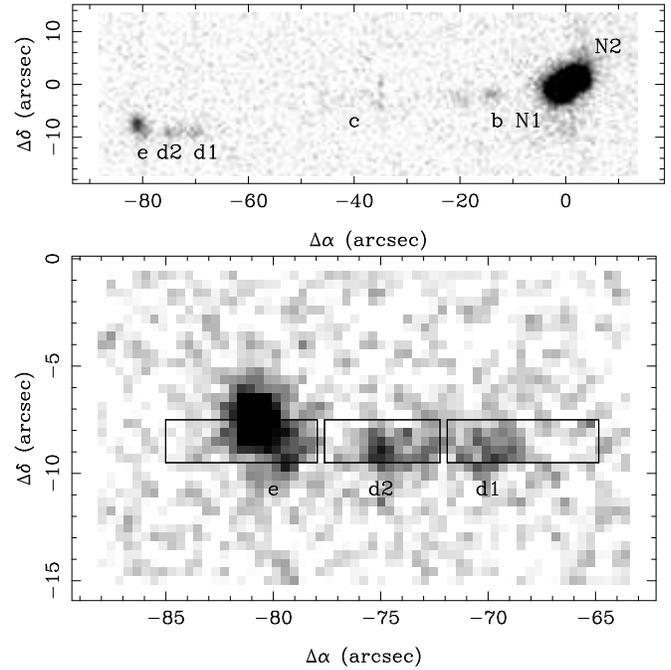

\includegraphics[width=33mm,angle=270]{1884_f2.ps}
\includegraphics[width=55mm,angle=270]{1884_f3.ps}
\caption{Continuum--subtracted \ion{H}{$\alpha$} image of Mkn 298 with overlaid the 
position of the ESO slit on the d,e knots.\label{fig:haim}}
\end{figure}

The positions of the slits and MPFS spectra are displayed in Figs. \ref{fig:bim} and \ref{fig:haim}, where they are overplotted to the R-band and 
\ion{H}{$\alpha$} images of Mkn 298 respectively.
The \ion{H}{$\alpha$} image clearly shows the existence of two knots in the
nucleus which were further resolved in four knots (CK,CKN1 and CKN2,CKN4) 
by \citet{moles}. 
The knots are not visible in the $R$-band image.

A $2\arcsec \times 2\arcsec$ rebinning of the MPFS spectra was done 
before the measurements in order to increase the signal to noise 
ratio: each measured spectrum is therefore the sum of four MPFS spectra. 
Spectra from R2 were used to obtain a 2D map
of the emission line ratios with the highest spectral resolution.
Finally, two regions corresponding to the 
nuclear knots (N1, N2) were defined and spectra inside them were summed 
together. In this case, spectra from R1 were used to cover the full 
spectral range from \ion{[O}{ii]}$\lambda3727$ to 
\ion{[S}{ii]}$\lambda6716,31$, but with a lower spectral resolution.

Emission line ratios after corrections for reddening are shown in 
Table~\ref{tab:ratios}. 
For each region, we selected the spectra with the best spectral 
resolution, namely the MPFS spectra for the nucleus and the ESO spectra for 
the d,e knots. Uncertainties on the measured line ratios are $\le$ 10\% in the nucleus, 
$\sim$ 20-30\% in the knots.
The uncertainty is much higher ($\sim$ 50\%) in the case of \ion{[O}{iii]}$\lambda$4363,
due both to the nearby \ion{H}{$\gamma$} $\lambda$4340 and to the noise
in the continuum.

The redshift of Mkn 298 as derived from the peak positions of the lines
is $z = 0.034$, giving a linear scale 0.66 kpc/arcsec ($H_0$ = 75 km s$^{-1}$
Mpc$^{-1}$).

\begin{figure}
\includegraphics[width=85mm,clip=]{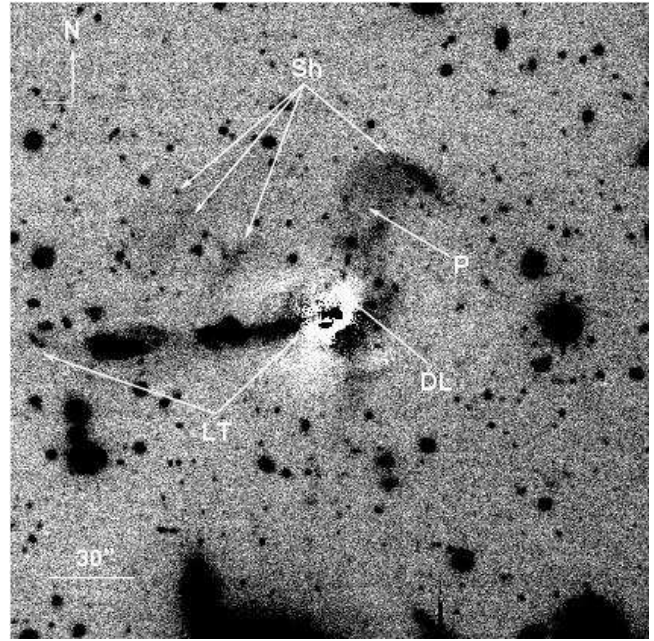}
\caption{Residuals after the GIM2D two-dimensional model subtraction: white 
arrows indicate the long tail (LT), the dust lane (DL), the plume (P) and the 
shells (Sh).}
\label{res}
\end{figure}

\begin{figure}
{\hbox{\includegraphics[width=85mm]{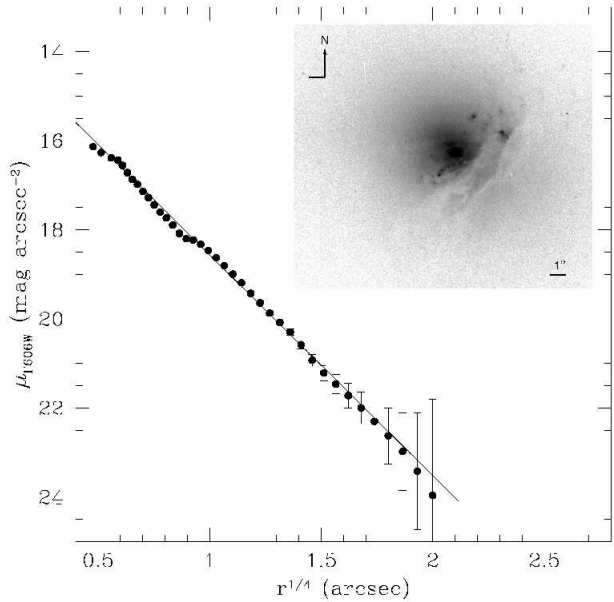}
}}
\caption{HST image of Mkn 298 and fit of the surface 
brightness profile.\label{fig:hst}}
\end{figure}

\section{Morphology\label{sec:morphology}}

Understanding the peculiar morphology of Mkn 298 requires a detailed 
analysis of deep images. 
For this purpose we applied to the NTT R-band image the GIM2D 
package \citep{simard}, which performs a two-dimensional
bulge/disk decomposition of galaxy images, and produces as output the model 
convolved with the PSF and the residual image (= galaxy $-$ model).
Mkn 298 was first fitted using a Sersic $R^{1/n}$-law for the bulge and a
classical exponential law for the disk. The fit gave as result an 
index $n=3.5$ and a reduced-$\chi^2\sim1.28$.
Then  a deVaucouleurs $R^{1/4}$-law was applied for the bulge obtaining 
quite similar residuals (reduced-$\chi^2\sim1.16$).
The resulting B/T ratio $\sim$ 0.78 indicates clearly that Mkn 298 is a 
spheroidal galaxy, likely a S0, with a weak disk inclined at $i \sim 40\degr$. 
The fit gave also the effective radius of the bulge, $r_e \sim 7$\arcsec\ and the 
radius $r_{half} \sim 9.5$\arcsec\ containing half of the total integrated light.

The structure and morphology of the ionized gas in the nucleus was discussed
in detail by \citet{moles}. In addition, we note that 
the strong nuclear  dust lane, whose maximum radial extension is  
$\sim 17$\arcsec ($\sim 11$ kpc), leaves a residual resembling a ring inclined 
at $i \sim 140-145\degr$. 
Interestingly, the galaxy shows at least three shells located 
at $\sim$ 40\arcsec, 1\arcmin~ and 1.3\arcmin ($\sim$ 26, 42 and 50 kpc
respectively) at East, North-East to the nucleus, and another one 
at $\sim$ 58\arcsec~($\sim$ 38 kpc) North-West to the nucleus, connected to a
large plume oriented at P.A. $\sim 160$\degr, and named Second Tail by 
\citet{moles} (see Figure~\ref{res}). 
These features are not uncommon in
elliptical and S0 galaxies, and are still studied to understand their real
nature and origin. The simulations by \citet{hernquist} 
show that mergers between equal-mass galaxies can
develop such shells on timescales of about 1-2 Gyr. \citet{turnbull} studied 
two 
examples of shell ellipticals, which have very strong analogies with those 
present in Mkn 298 (see their Figs. 9 and 10). 
Nevertheless, Mkn 298 is a rare example of early-type galaxy showing 
both shells
and a prominent stellar and gaseous tidal tail, a situation not seen
in the above mentioned numerical simulations.
Faint tidal tails were found by \citet{bal97} in the shell elliptical 
NGC 3656,
and several shells and a long H\,{\sc i} tidal tail were discovered by
\citet{vm02} in HCG 54.
Only the NW shell has a surface brightness close to 3$\sigma$ above the
background. Assuming a sky brightness $\mu_R \sim$ 20.5 mag arcsec$^{-2}$, 
that is the value expected at LaSilla for similar moon illumination 
conditions, we obtain $\mu_R \sim$ 25 mag arcsec$^{-2}$: such a value is in 
agreement with those measured by Turnbull et al. for the shells of their 
galaxies.
The other shells in Mkn 298 have surface brightness between 1 and 2$\sigma$ 
of the background.
In the residual image it can be also seen  that the Eastern tidal tail extends
beyond the d and e knots, up to 1.75\arcmin ($\sim 70$ kpc),  where it 
seems to  turn back toward  North-West.

The nucleus of Mkn 298 was also observed by HST with the WFPC2 and the 
broad-band $V$ filter  F606W \citep{malkan}. 
An analysis of the   isophotes of the galaxy was done using the 
ELLIPSE task in IRAF.
Since the presence of the dust lane makes difficult to fit the real 
isophotes, we applied the masking method used by \citet{carollo}. 
The surface brightness profile plotted against $r^{1/4}$ 
can be fitted by a de Vaucouleurs law ($r_e \sim 8$\arcsec) without an 
additional disk
component; it does not show any nuclear point-like source (e.g. AGN or stellar
cluster), in agreement with the results given by GIM2D.

In summary, the brightest part of Mkn 298, that is within $\sim 6.3$ kpc, is
clearly spheroidal. Such morphology, combined with the presence of a dust 
lane,  a bright and extended tail and some weak shells, are
all strong clues in favor of the hypothesis that Mkn 298 has experienced a 
major merger event.

\begin{table}
\caption{\ion{H}{$\alpha$} luminosities and star formation rates. 
The labels in parentheses are those adopted by \citet{moles}.}
\label{tab:ha}
\begin{tabular}{lcc c cc}
\hline\hline
 & \multicolumn{2}{c}{Observed} & & \multicolumn{2}{c}{Dereddened}\\
& Log $L_{\rm H\alpha}$  &  SFR & & Log $L_{\rm H\alpha}$  &  SFR \\
\hline
N1 (CK+CKN1) & 41.46 & 2.30 & & 41.83 & 5.35 \\
N2 (CKN2+CKN4)& 41.21 & 1.29 & & 41.50 & 2.52 \\
a (galaxy body) & 41.72 & 4.19 & & 42.01 & 8.14 \\
b (MTK1) & 40.02 & 0.08 & & & \\
c (MTK3) & $\le$ 39.87 & $\le$ 0.06 & & & \\
d1 (MTK4) & 39.93 & 0.07 & & 40.52 & 0.26 \\
d2 (MTK5) & 40.03 & 0.08 & & 40.58 & 0.30 \\
e (MTK6) & 40.42 & 0.21 & & 40.42 & 0.21 \\
\hline
\end{tabular}
\end{table}

\begin{figure*}
\hbox{
	\includegraphics[width=80mm]{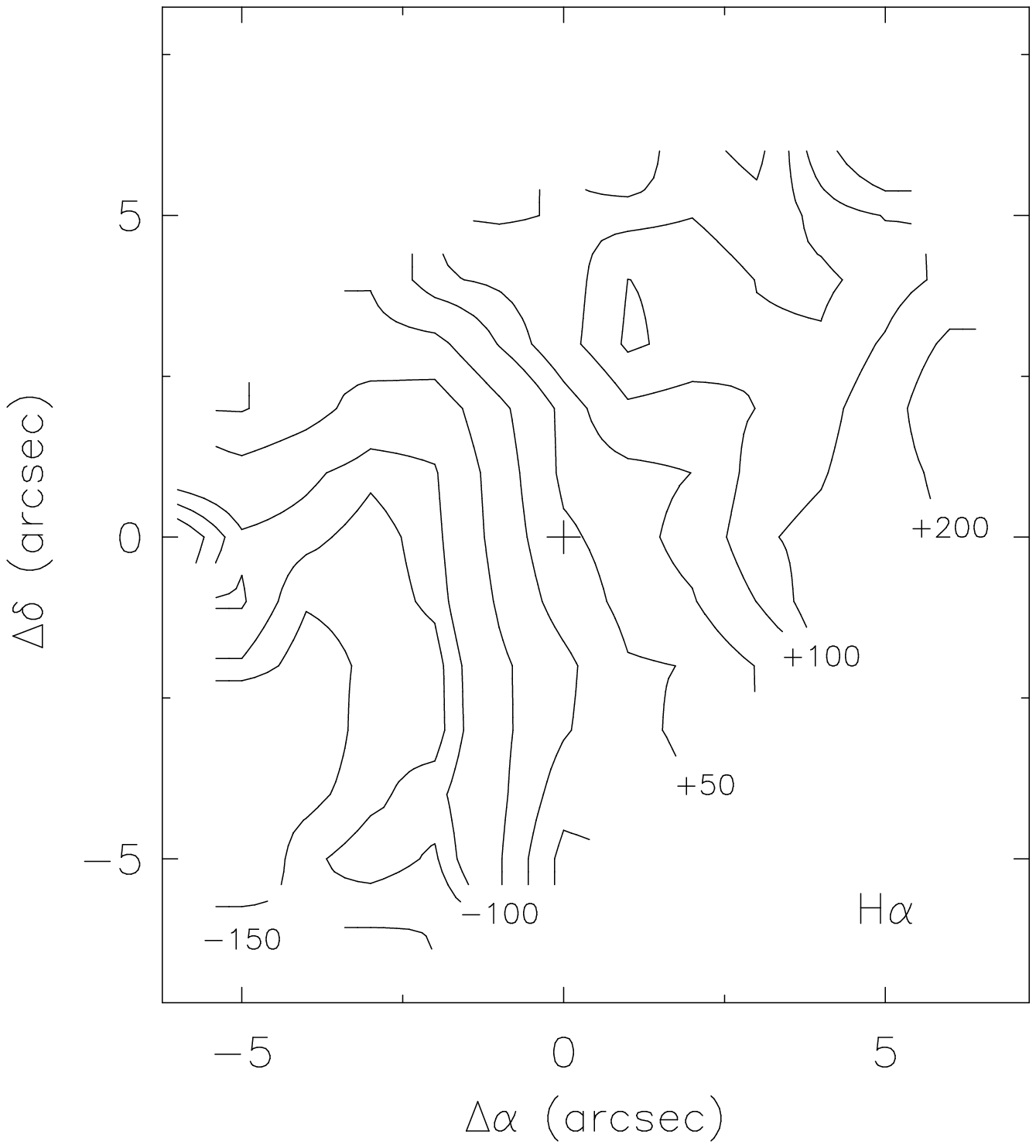}
	\includegraphics[width=85mm]{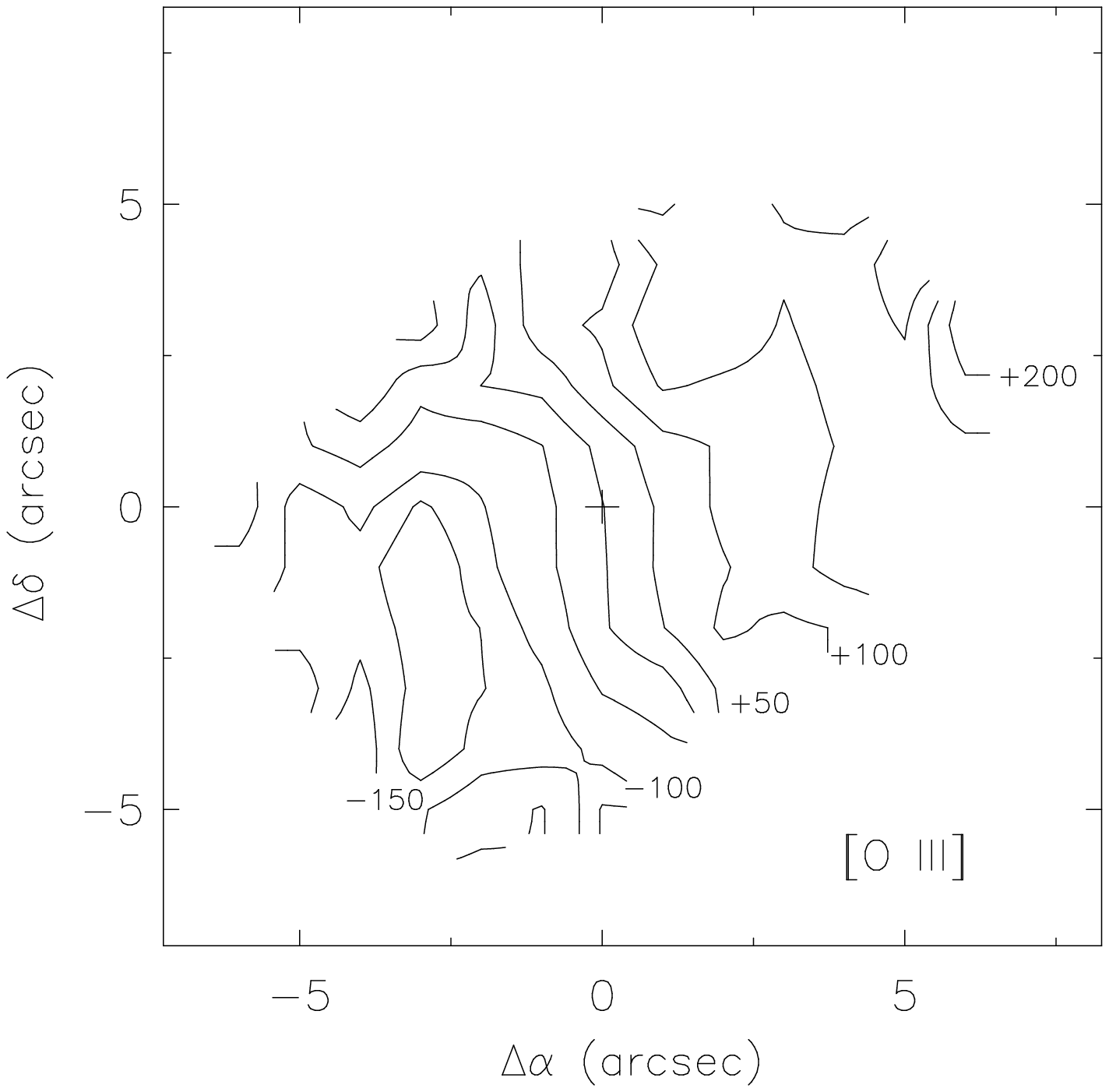}
}
\hbox{
	\includegraphics[width=80mm]{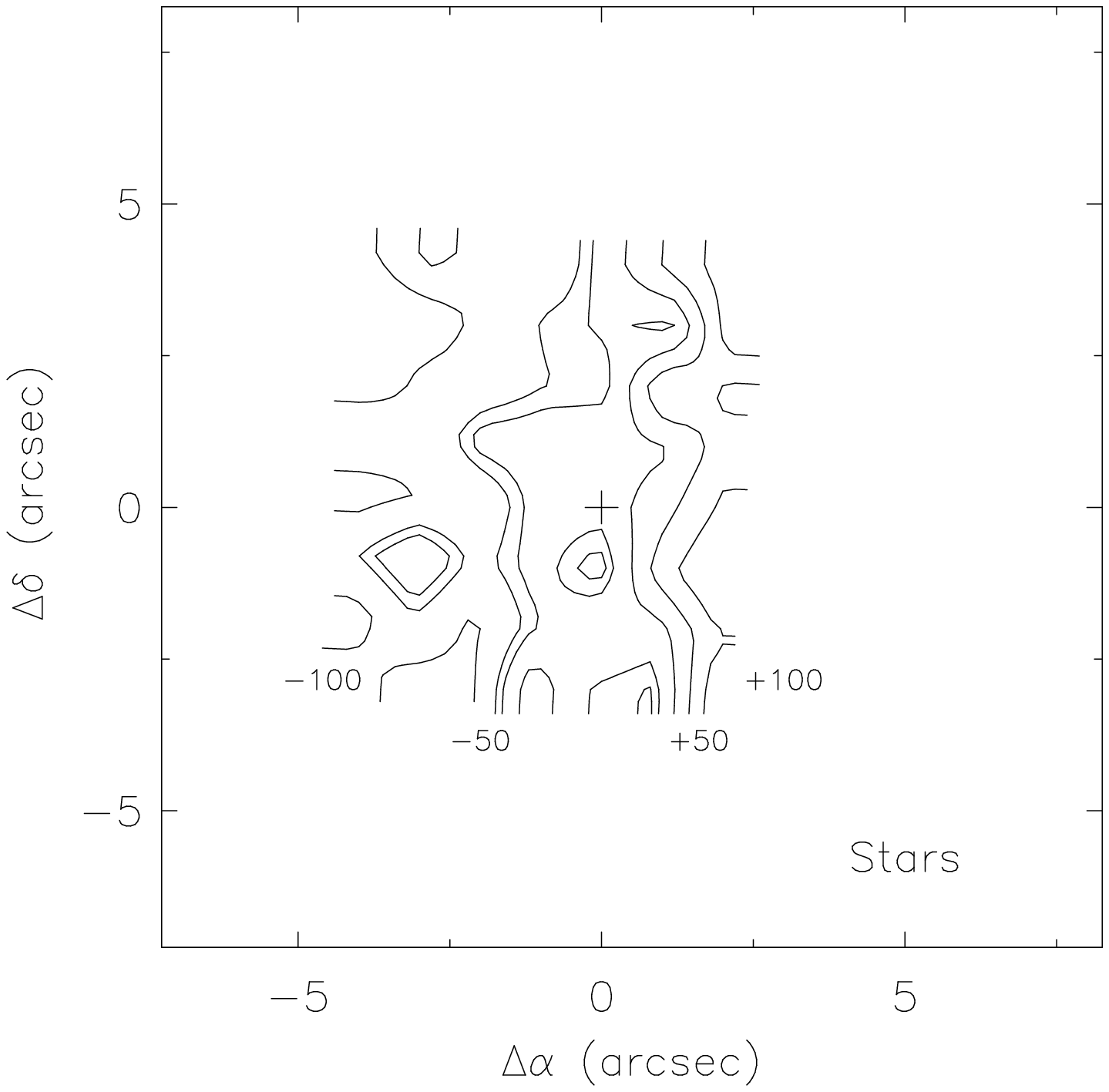}
	\hskip 17pt \includegraphics[width=80mm]{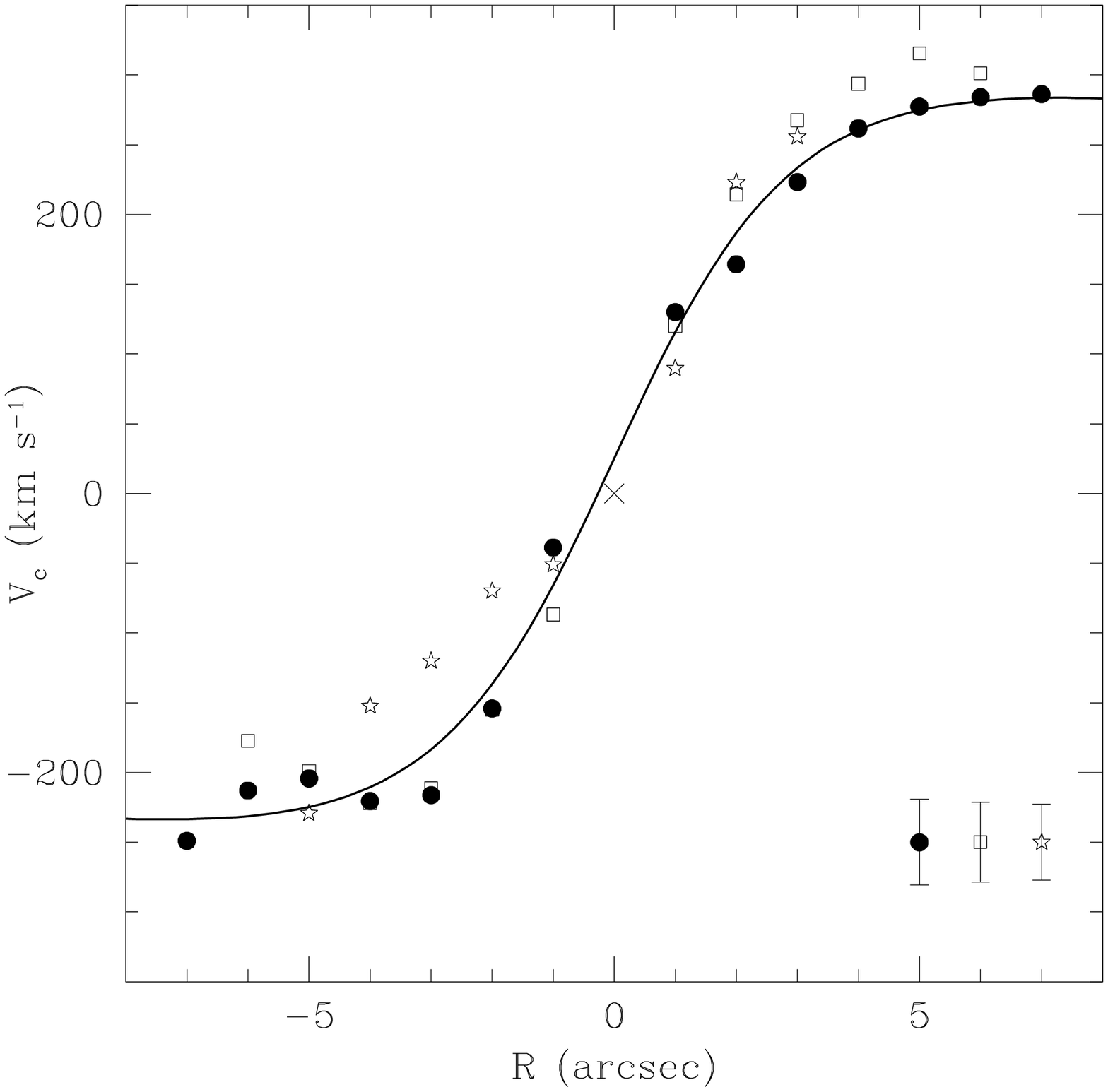}
}
\caption{Velocity fields of \ion{H}{$\alpha$} (up-left), 
\ion{[O}{iii]}$\lambda$5007 (up-right), and stars (bottom-left); the cross
indicates the position of the nuclear continuum. North is up and East on the
left. (bottom-right) The deprojected velocity curves of \ion{H}{$\alpha$}
(filled circles), \ion{[O}{iii]}$\lambda$5007 (open squares) and stars (open
stars). The solid line is the pure circular motion fit of the gaseous component.}
\label{kin}
\end{figure*}

\section{Star formation rates\label{sec:sfr}}

The presence of multiple knots in the nucleus is also visible in the continuum
subtracted \ion{H}{$\alpha$} image. Aperture photometry was done on the two brightest
nuclear knots (N1, N2) and on the regions in the tail (b,c,d,e). A very faint 
H$\alpha$ emission is detected in the c knot. 

An estimate of the photometric accuracy was done by comparing
the so-obtained fluxes  with those obtained from the MPFS 
spectra. We conclude that the photometry of imaging and spectroscopic data is 
consistent within 5\%.

\ion{H}{$\alpha$} 
luminosities obtained in each of these regions are shown in 
Table~\ref{tab:ha}; for those regions where dust extinction from 
\ion{H}{$\alpha$}/\ion{H}{$\beta$} was derived from spectra, the extinction corrected 
values are also
displayed. \ion{H}{$\alpha$} luminosities were finally converted to star 
formation rates using Equation 2 from \citet{kennicutt}. 
In the nucleus, this is an upper limit if part of the \ion{H}{$\alpha$} 
emission  is due to an AGN. 
We see that  the total star formation rate  ($a$ region)
 is $\sim$ 8.1 M$_{\sun}$ yr$^{-1}$, well below 
the value of 90 M$_{\sun}$ yr$^{-1}$ given by \citet{moles}. 
The \ion{H}{$\alpha$} luminosities in the galaxy measured by us 
and by \citet{moles} before correction for dust reddening are consistent 
within 30\%. The difference in the SFR is  mainly due to the internal 
reddening adopted by \citet{moles}, $E(B-V) \sim 1$, which is much higher 
than the average reddening that we find in the nucleus, $E(B-V) \sim 0.3$.
This  value of the reddening is also in agreement with the value found by
\citet{koski}, $E(B-V) = 0.40$. A possible reason for this discrepancy 
may be that the spectra that \citet{moles} used to derive the extinction
come from knots (CK and CKN2) which are very close to the dust lane: they
could be therefore biased to high values of 
\ion{H}{$\alpha$}/\ion{H}{$\beta$}, not representative of the whole 
galaxy.

Mkn 298 is not a bright IRAS source; the IRAS Faint Source Catalog
contains an entry which is however centered on the d,e knots.
Approximate fluxes at the position of Mkn 298 can  
be derived from the Iras Sky Survey Atlas using the  
{\em IRSA}\footnote{NASA/IPAC Infrared Science Archive}
facility: we obtain $f_{60\mu m} = 0.8$ Jy,
 $f_{100\mu m} = 2.1$ Jy.
 These values are very uncertain as the emission is distributed 
on a wide area ($>$ 2 arcmin around the galaxy) and is probably
the sum of the emission from the galaxy and from the tidal tail.
According to Equations 7, 8 from \citet{hopkins}, 
we obtain $L(8-1000\mu m) = 1.8\times10^{44}$ erg s$^{-1}$ and a 
SFR $\sim 11$ M$_{\sun}$ yr$^{-1}$, a value in good agreement with the
SFR derived from the total \ion{H}{$\alpha$} luminosity 
(galaxy+tidal tail). As the contribute to the SFR from the tidal
tail estimated from the \ion{H}{$\alpha$} luminosity is negligible
compared to that in the galaxy, it seems likely that this is also the case
for the infrared luminosity.

\citet{geo} explored the possibility to use
the $L_{\rm FIR}/M(H_2)$ ratio (star formation efficiency, SFE hereafter) 
as an indicator of the galaxy merging age. In the case of Mkn 298, we obtain
SFE $\le 9$ L$_{\sun}$/M$_{\sun}$ using the above far--infrared luminosity 
and the value   $M_{\rm mol} \sim 5 \times 10^9$ $M_{\sun}$ from \citet{braine}. 
Together with the low fraction of HI in the nucleus,
this value would indicate that Mkn 298 is in a post-merger stage. The N1 and 
N2 knots in the nucleus are probably separate star forming regions 
partially  obscured by the dust lane, rather
than separate nuclei before merging.

\begin{figure*}
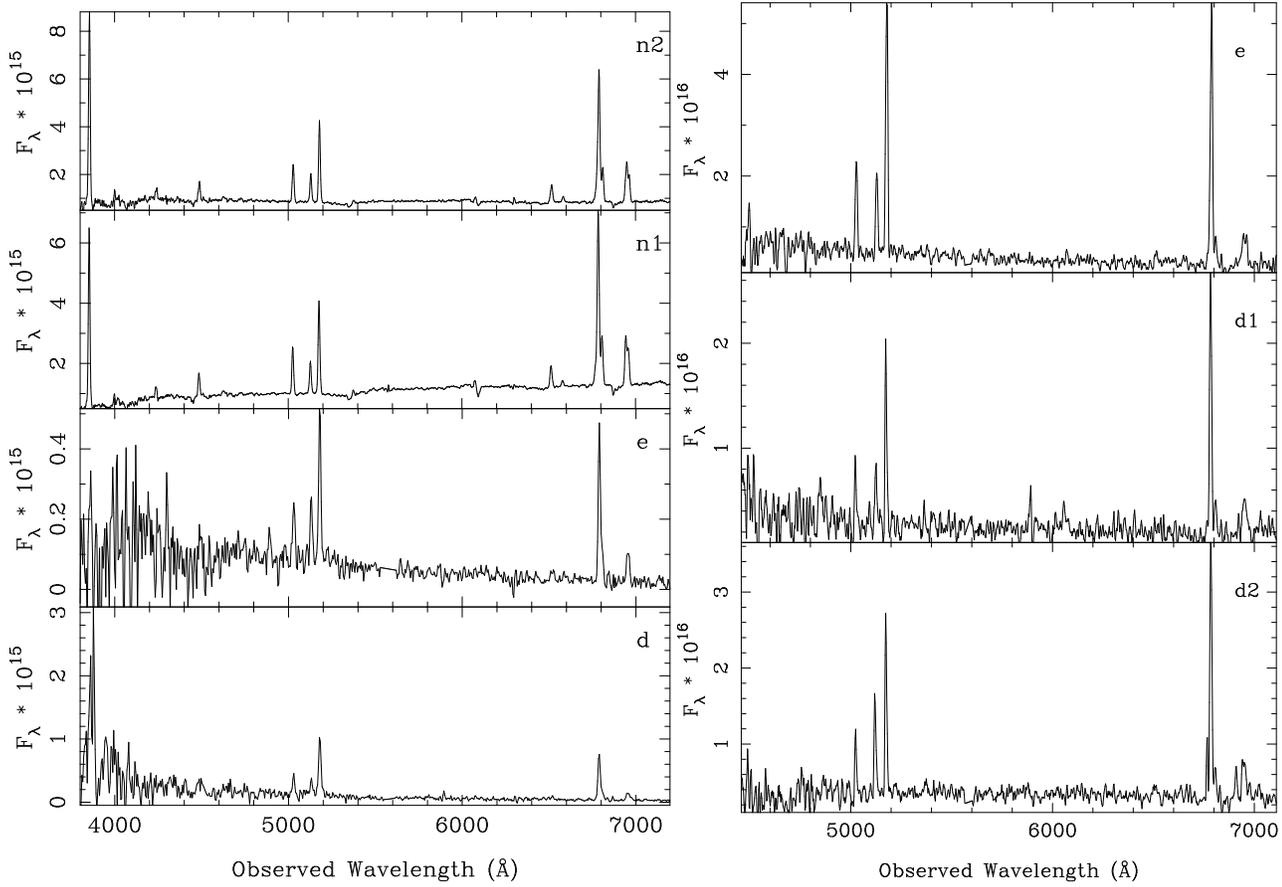

\includegraphics[width=88mm]{1884_f10.ps}
\includegraphics[width=80mm]{1884_f11.ps}
\caption {1D spectra extracted in the nucleus and in the d,e knots from the 
MPFS  ({\em left}) and EFOSC 2 ({\em right}) data. }
\end{figure*}

\begin{table*}
\caption{Dereddened emission line ratios (\ion{H}{$\beta$}=1.00)
and models 
computed assuming (a) photoionization only and (b) photoionization
+ shock (see text for model parameters). The values in italics are the 
uncertainties on the observed line ratios; the 
\ion{H}{$\alpha$}/\ion{H}{$\beta$} ratio before reddening correction is also given below.}
\label{tab:ratios}
\begin{tabular}{lccccc ccccccc}
\hline\hline 
& MPFS & MPFS& ESO & ESO & ESO 
& \multicolumn{2}{c}{Cloudy} &
 \multicolumn{2}{c}{AGN + shock} &
 \multicolumn{3}{c}{SB + shock} \\
& N1 & N2 & d1 & d2 & e & Nucleus & d,e & N1 & N2 & N1 & N2 & d,e \\
\hline
\ion{[O}{ii]} $\lambda$3727 & 5.3 & 6.5 & & & 3$^1$ & 6.98 & 3.92 & 6.00 &
4.90 & 5.30 & 5.70 & 3 \\
    &{\em 0.1 }  & {\em 0.3} & & & {\em 1}  \\

\ion{[Ne}{iii]} $\lambda$3869 & 0.20 & 0.2 &  &  &  & 0.64 & 0.24 & 0.20 & 0.20 &
0.28 & 0.28 & 0.57 \\
    &{\em 0.04 }  & {\em 0.1} & & &   \\

\ion{H}{$\gamma$} $\lambda$4340 & 0.56 & 0.65 &  & & & 0.47 & 0.47 & 0.44 & 0.44 &
0.46 & 0.46 & 0.46 \\ 
    &{\em 0.04 }  & {\em 0.05} & & &   \\

\ion{[O}{iii]} $\lambda$4363 & 0.03: & 0.08: &  &  & & 0.02 & 0.02 & 0.19 & 0.16 &
0.03 & 0.02 & 0.02 \\ 
    &{\em 0.02 }  & {\em 0.03} & & &   \\


\ion{[O}{iii]} $\lambda$4959 & 0.67 & 0.70 & 0.7 & 1.9 & 0.8 & 0.95 & 1.01 & 0.58
& 0.50 & 0.67 & 0.54 & 0.95 \\
    &{\em 0.03 }  & {\em 0.04} &{\em 0.5} &{\em 0.5} & {\em 0.1}  \\

\ion{[O}{iii]} $\lambda$5007 & 1.90 & 1.98 & 3 & 2.9 & 2.7 & 2.85 & 3.02 & 1.73
& 1.50 & 2.00 & 1.62 & 2.85 \\ 
    &{\em 0.06 }  & {\em 0.08} &{\em 1} &{\em 0.7} & {\em 0.3}  \\

\ion{[O}{i]} $\lambda$6300 & 0.35 & 0.40 & $<$0.1 & $<$0.1 & $\le 0.1$  & 0.46 &
0.06 & 0.30 & 0.33 & 0.30 & 0.35 & 0.14 \\
    &{\em 0.02 }  & {\em 0.02} & & &   \\

\ion{H}{$\alpha$} $\lambda$6563 & 2.80 & 2.80 & 2.80 & 2.80 & 2.80 & 2.91 & 2.87 & 3.30 &
3.40 & 2.97 & 3.00 & 2.93 \\ 

\ion{[N}{ii]} $\lambda$6583 & 0.86 & 0.65 & 0.4 & 0.3 & 0.22 & 0.92 & 0.23 & 0.80
& 0.90 & 0.82 & 0.93 & 0.33  \\
    &{\em 0.03 }  & {\em 0.03} & {\em 0.2}&{\em 0.1} & {\em 0.05}  \\

\ion{[S}{ii]} $\lambda$6716 & 0.72 & 0.83 & & & 0.32 & 1.24 & 0.30 & 0.80 & 0.86 &
0.72 & 0.81 & 0.37 \\
    &{\em 0.03 }  & {\em 0.03} &{\em } &{\em } & {\em 0.05}  \\

\ion{[S}{ii]} $\lambda$6731 & 0.54 & 0.51 &  &  & 0.30 & 0.92 & 0.40 & 0.67 & 0.82 &
0.60 & 0.76 & 0.29 \\
    &{\em 0.02 }  & {\em 0.02} &{\em } &{\em } & {\em 0.05}  \\

\ion{[S}{ii]} $\lambda$6716+31 & 1.26 & 1.34 & 0.9 & 0.7 & 0.6 & 2.16 & 0.70 &
1.47 & 1.68 & 1.32 & 1.57 & 0.66 \\ 
    &{\em 0.05 }  & {\em 0.05} & {\em 0.3} &{\em 0.2} & {\em 0.1}  \\

\hline

\ion{H}{$\alpha$} $\lambda$6563 & 4.3 & 4.2 & 6 & 5 & 2.8 \\ 
    &{\em 0.1 }  & {\em 0.1} & {\em 2} & {\em 1} & {\em 0.2}  \\

E(B-V)& 0.40 & 0.37 & 0.61 & 0.57 & 0.00 & \\
\hline
$^1$ from MPFS data 
\end{tabular}
\end{table*}

\section{Kinematics\label{sec:kinematics}}

Taking advantage of the integral field data, we carried out a two-dimensional 
investigation of the gaseous and stellar kinematics in Mkn 298.

First, the wavelength positions of the two brightest emission lines 
\ion{[O}{iii]}$\lambda$5007 and \ion{H}{$\alpha$} were measured by means of gaussian fitting 
of their profiles in each spectrum where they could be detected down to 
3$\sigma$ over the continuum.
Then, the Mg$b$ $\lambda$5175 absorption triplet was chosen for deriving stellar
kinematics, and the Fourier cross-correlation method was applied by means of the 
IRAF task FXCOR \citep[][ TD79 hereafter]{td79}, and using a mean sky-flat 
spectrum as template. 
Only the correlation peaks with the TD79 parameter R $> 5$, which is an
indicator of the signal-to-noise ratio, were considered reliable. 
The so obtained radial velocities were corrected for the motion of the observer 
in the direction of the observation, and then re-organized to construct the
velocity fields for each component (Figs. \ref{kin}).           
A first look reveals that the \ion{[O}{iii]} and \ion{H}{$\alpha$} kinematics are in agreement,
showing similar displacement of their ``spider'' diagrams, which are extended up
to a maximum radius of $\sim 7\arcsec$ ($\sim 4.6$ kpc), and concordant
velocity values. In addition, their velocity fields have the major axes clearly
parallel to the dust lane, strongly suggesting that the gas and the dust lane
belong to the same structure, a rotating disk of a likely former spiral galaxy
merged into a early-type, or however a poor-gas galaxy.
On the contrary, stars show a velocity field with a slightly 
different orientation, indicating the presence of a kinematical decoupling.

In order to determine the rotation curve of the galaxy in the field of view of
the spectrograph, we applied the GIPSY (Groningen Image Processing SYstem) 
task ROTCUR. Based on the method
described by \citet{begeman}, ROTCUR derives kinematical parameters by fitting 
the velocity field with concentric tilted-rings, characterized by seven free
parameters.  In our case, given the relatively small number of available
measurements, it was necessary to conveniently reduce the degrees of freedom.
We followed the recipes by \citet{begeman} to determine the $X_0$ and $Y_0$
coordinates of the center, and the systemic velocity, $V_{sys} \sim 10250$ 
\kms, and in addition we fixed the inclination, $i$, of the rings on the 
basis of the optical isophotes. Therefore, ROTCUR fitted the circular velocity,
$V_c$, for increasing radii, varying the position angles of the rings.
We performed this procedure separately for the approaching and the receding
sides of the velocity field, in order to account for possible distortions or
asymmetries.
The resulting $V_c(R)$ values were plotted to visualize the deprojected rotation
curve for gas and stars. As mentioned above, \ion{[O}{iii]} and \ion{H}{$\alpha$} show similar
radial velocities, almost symmetrically distributed, with a $\Delta V_{max} \sim
250$ \kms. Stars are well in agreement with the gaseous component 
only in the receding side, while poorly in the approaching side, with lower
velocities up to -100 \kms with respect to the gas. 
From the analysis of the position angles of the rings, we obtained that the 
kinematical axes of gas and stars are not aligned, as expected, and oriented at
$\sim 30\degr$ and $\sim 0\degr$ respectively. Moreover, the kinematical axis 
of the gas is more stable in the receding side than in the approaching one, 
where the variations can reach 20\degr.

We have compared our rotation curve to that published by \citet{moles}, 
finding a general good agreement for the gaseous component. These authors used 
a long-slit spectrum taken close to the major axis, and obtained a curve with 
similar radial extension and velocities. Thanks to their higher spatial 
resolution they could claim about two rotating systems, which unfortunately are
not visible with our data. On the contrary, we can assert that the stellar
component shows rotation, at least in the region of the galaxy where the stellar
kinematics could be measured.

We fitted the \ion{H}{$\alpha$} curve with an analytic relation, which assumes a
spherical potential model with pure circular motions, 
$V_c(R)=aR/(R^2+c_0^2)^{p/2}$ \citep{ber91}.

The nonlinear least-squares fit was stable, giving the following parameters:
$a=515$ \kms, $c_0=3.8$\arcsec and $p=1.27$. Using this relation, we
estimated the total mass within a radius of 7\arcsec 
obtaining $\sim 7.3\times10^{10}$ M$_{{\sun}}$.

The nuclear stellar velocity dispersion, $\sigma_0$, was also measured.
First, the brightest central spectrum was considered, and then an average of 
central spectra within an aperture of 3\arcsec. The results were similar.
The estimate of $\sigma_0$ was obtained again with FXCOR by measuring the FWHM
of the correlation peak, and then converting into $\sigma_0$ following the
recipes by \citet{nw95}. In detail, we convolved the template
spectrum with gaussian functions of increasing known $\sigma$, and we
cross-correlated each convolved spectrum with the original one, obtaining a sort
of $\sigma$-FWHM empirical relation, which becomes flat for $\sigma < 90$ \kms.
The nuclear $\sigma_0$ resulted $\sim 170$ \kms, a value in 
agreement with the one measured by \citet{moles}, even if slightly lower.

Given the existence of a Mg-$\sigma$ relation in early-type galaxies and spiral
bulges, we measured the LICK/IDS system Mg and Fe indices in the center of the 
galaxy: $Mg_1,Mg_2, Mg~b, Fe5270, Fe5335$, defined by \citet{w94}, and we 
attempted a comparison with published values and relations. 
Since emission lines are sometimes present at the blue or red edges of
these absorption features, the estimate of the continuum may be rather 
uncertain. Therefore, our measurements should be considered as rough 
estimates.
Anyway, $Mg_1=0.08$, $Mg_2=0.17$, $Mg~b=3.6$ and
$<Fe>=0.5(Fe5270+Fe5335)=2.07$ are values 
typically observed in spiral bulges, and not in ellipticals or S0s \citep[see
e.g.][]{ps02} at the $\sigma_0$ measured in Mkn 298. 
In other words, Mkn 298 is an outlier of the $\sigma$-Mg and
$\sigma$-Fe relations, showing indices with lower strength  
with respect to spheroidal galaxies. This effect can be caused either by age 
or by metallicity.
On the other hand, the [Mg/Fe] ratio is
around 0.4, indicating that the bulk of the nuclear stellar component is old
probably formed in a timescale of 1-2 Gyr \citep{id96}. 
We can conclude that Mkn 298 does not fit the properties of the early-type
galaxies because of the ongoing merger. Indeed, bursts of star 
formation induced by the strong interaction can effectively make the Mg and Fe 
indices strengths weaker because of the presence of bright, hot
stars \citep{wc03}. 

Similar results were obtained by \citet{lon00}, who showed that shell-
galaxies, believed to be the product of a past strong interaction, deviate
 significantly from the $\sigma$-Mg and $\sigma$-Fe relations (see their 
Fig. 2).  They proposed that this may be due  either to  the  
gravitational interaction which affects the velocity 
dispersion or to  bursts of star formation that alter the metallicity 
indices, or to a combination of both.



\begin{figure}
\includegraphics[width=70mm,angle=270]{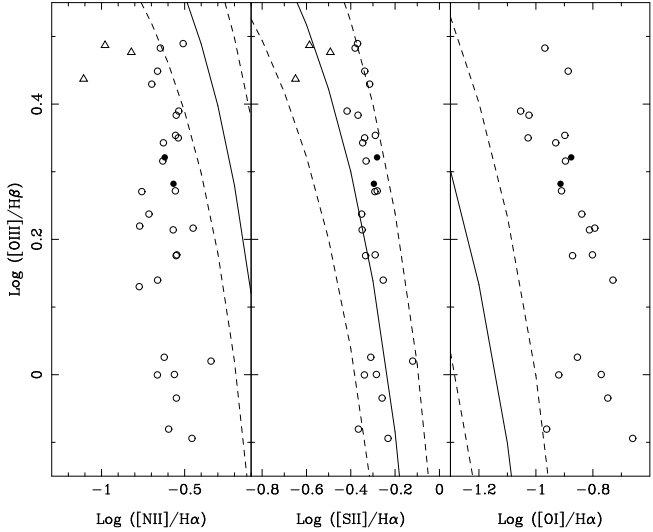}

\caption {VO diagnostic line ratios in Mkn 298. 
{\em Circles}: binned MPFS spectra in the nucleus.
{\em Dots}: integrated MPFS spectra in the nucleus (N1, N2).
{\em Triangles}: ESO spectra in the d,e knots.  
 The solid line
is the theoretical starburst border from Kewley et al. (2001), dashed lines
are $\pm 0.1$ dex of these lines. \label{fig:kewley}}
\end{figure}

\begin{table}
\caption{Statistics of dereddened emission line ratios in the nucleus 
from the MPFS data (R1).}
\label{tab:ratios2d}
\begin{tabular}{l c c c c}
\hline\hline
& min & max & avg & $\sigma$ \\
\hline
\ion{[O}{ii]} $\lambda$3727     &    3.53 &    9.78  &    7.39 &   1.89 \\
\ion{[O}{iii]} $\lambda$5007    &    0.78 &    2.90  &    1.69 &   0.61 \\
\ion{[O}{i]} $\lambda$6300      &    0.27 &    0.63  &    0.40 &   0.09 \\
\ion{H}{$\alpha$} $\lambda$6563  &    2.18 &    2.80  &    2.76 &   0.13 \\
\ion{[N}{ii]} $\lambda$6583     &    0.47 &    0.99  &    0.72 &   0.14 \\
\ion{[S}{ii]} $\lambda$6716     &    0.56 &    1.08  &    0.79 &   0.12 \\
\ion{[S}{ii]} $\lambda$6731     &    0.41 &    0.63  &    0.51 &   0.06 \\
\ion{[S}{ii]} $\lambda$6716+31  &    1.01 &    1.65  &    1.30 &   0.16 \\
                             &         &          &         & \\
                   E(B-V)    &    0.03 &    1.58  &    0.50 &   0.42 \\
\hline
\end{tabular}
\end{table}

\section{Emission line ratios\label{sec:linerat}}

Figure~\ref{fig:kewley} shows the VO line ratios observed both in the nucleus
and in the d,e knots:  
overlaid are the lines dividing the theoretical starburst
region from other types of excitation computed by \citet{kewley}. 
While in the d,e knots the line ratios are consistent with those
expected in a starburst, the situation in the nucleus is more complex. 
In fact, in both the nuclear knots \ion{[N}{ii]} 
$\lambda$6583/H$\alpha < 0.3$ is typical of starburst regions, 
\ion{[S}{ii]} $\lambda\lambda$  6716,6731 $\sim 0.5$
may be fitted by starburst models, whereas \ion{[O}{i]}/H$\alpha > 0.1$ is 
too high for any starburst model. 
A similar behavior may be seen in all the 
individual spectra taken by the MPFS array in the nucleus. 
Figure~\ref{fig:maps} and Table~\ref{tab:ratios2d} show that 
line ratios are remarkably constant over the whole galaxy (up to $5\arcsec$
from the center). Reddening by dust is low ($E(B-V) < 0.50$), with the 
exception of the regions on the western side  ($E(B-V) > 1$), where it is
due to the foreground dust lane.

\subsection{Chemical abundances}

As discussed in more detail in the next section, emission line ratios in the 
nucleus may imply the presence of an active nucleus. 
Chemical abundances in the nuclear regions were therefore 
computed using both the two calibrations 
proposed by \citet[][SSCK hereafter]{sb98} for the NLR of active galaxies
and the $R_{23}$ method (\citet{mcgaugh}, \cite{kobulnicky})  and 
 $p$ method \citep{pilyugin1,pilyugin2} for abundances in \ion{H}{ii} regions 
of  normal  galaxies. We did not use \ion{[O}{iii]}$\lambda$4363 due to the
strong uncertainty in its measurement.
 
The SSCK calibrations are based on  
\ion{[N}{ii]}$\lambda\lambda$6548,6583/\ion{H}{$\alpha$} vs. \ion{[O}{iii]}$\lambda\lambda$
4959,5007/\ion{H}{$\beta$}  (Eq. 2 in SSCK)  
and \ion{[O}{ii]}$\lambda$3727/\ion{[O}{iii]}$\lambda\lambda$ 4959,5007 (Eq. 3 in SSCK). Following the SSCK prescription we adopted the average of 
the two values,  12 + $\log$(O/H) $\sim$ 8.7 (Z/Z$_{\sun} \sim$ 0.7,
assuming 12+$\log$(O/H)$_{\sun}$ = 8.87).
To check these results we used the intensities given in Table~5
from \citet{moles} and obtained consistent results,
12 + $\log$(O/H) $\sim$ 8.5 (Z/Z$_{\sun} \sim$ 0.6).

The $R_{23}$ and $p$ methods allow to compute 
oxygen abundances from  
\ion{[O}{iii]}$\lambda\lambda$ 4959,5007/\ion{[O}{ii]}$\lambda$3727. 
Both give as result a lower abundance, 12 + $\log$(O/H) 
$\sim$ 8.3 (Z/Z$_{\sun} \sim$ 0.3).

As it concerns the d,e knots, \citet{iglesias} found 
12 + $\log$(O/H) = 8.41 (Z/Z$_{\sun} \sim$  0.35)  
using the $p$-method. A similar value, 
12 + $\log$(O/H) = 8.38 (Z/Z$_{\sun} \sim$  0.33) is obtained from the 
emission line ratios that we measure in the e knot.

We therefore conclude that there is evidence for a sub-solar metallicity throughout the galaxy. 
The metallicity in the nucleus (Z/Z$_{\sun} \sim$ 0.5-0.7) may be 
somewhat higher than in the d,e knots 
(Z/Z$_{\sun} \sim$ 0.3). We find no evidence supporting the very low 
metallicity (Z/Z$_{\sun} \sim$ 0.06) claimed by \citet{moles}.

Starting from these chemical abundances, we explored different models trying 
to fit the observed   line spectra:
pure photoionization models relative to an AGN and to a starburst, and 
composite
models which account for both photoionization and shocks.
In all the following models, $Z/Z_{\sun}$ was initially set to the value 
found here and then varied to give the best fit of the observed line ratios.
The model which leads to the best fit of the line ratios will determine
the controversial nature of Mkn 298.

\subsection{Photoionization models}

As a first attempt to model the emission line ratios measured in the
different regions, we assumed that they are produced by photoionization
from either an active nucleus or stars (\ion{H}{ii} regions). To this aim, the
photoionization code {\sc Cloudy 94} \citep{fer98} was used.

\begin{figure*}
\includegraphics[width=165mm,angle=270]{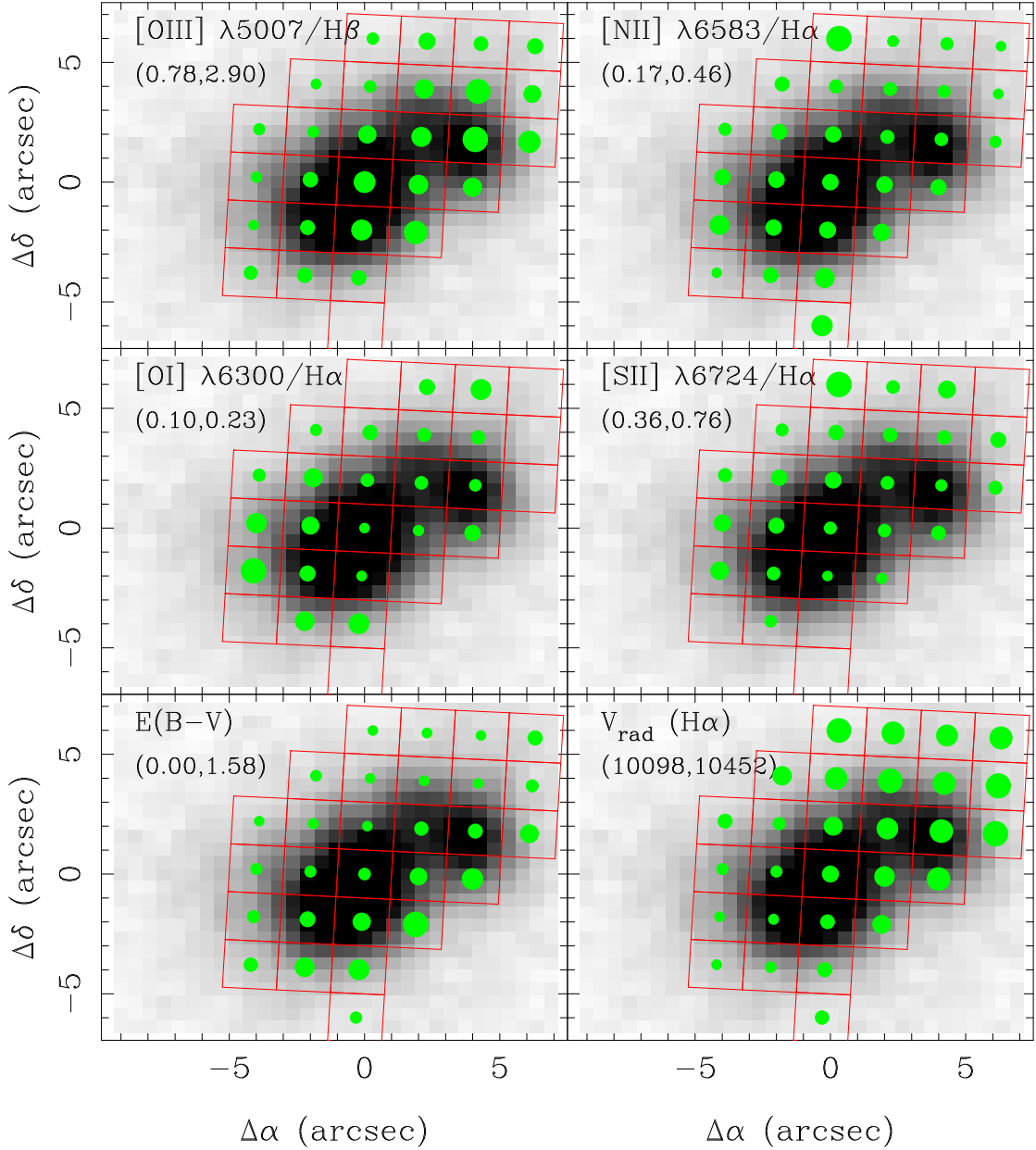}
\caption
  {Line ratios, dust extinction and \ion{H}{$\alpha$} velocities from 
the rebinned MPFS spectra (R2) in the nucleus are here displayed as circles
 whose size is
proportional to the measured values; the values in brackets are the minima 
and  maxima. Overlaid is the \ion{H}{$\alpha$} image. 
 \label{fig:maps} }

\end{figure*}

\begin{figure*}
\includegraphics[width=70mm,angle=270]{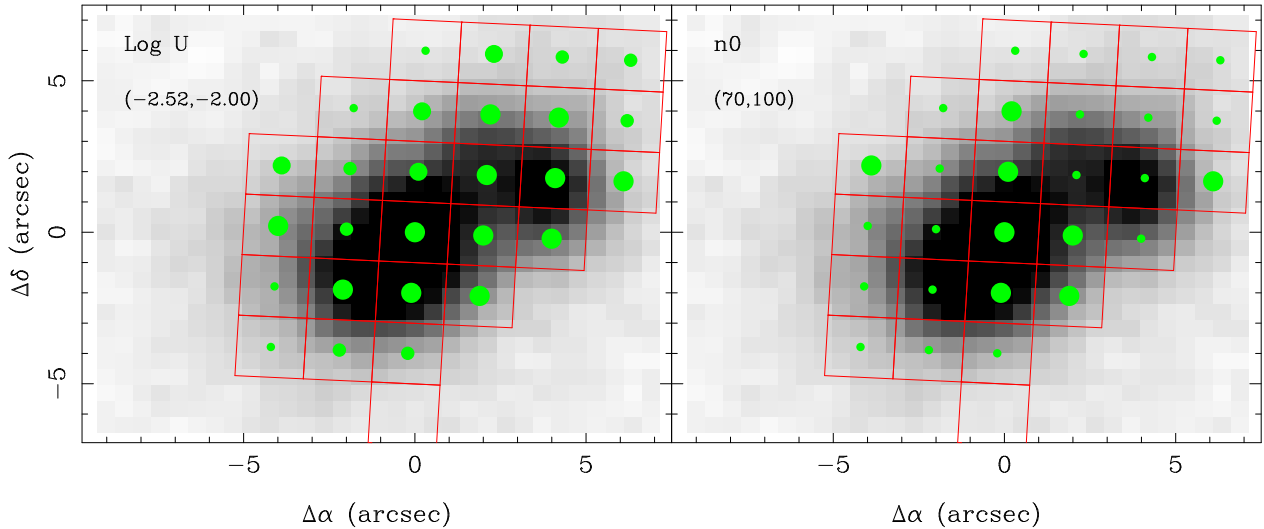}
\caption
 {Input parameters of the starburst+shock models which  
lead to the best fit of the  line ratios in Figure~\ref{fig:maps}: 
$\log U$ ({\em left}), \n0 ({\em right}).
 \label{fig:shmap} }

\end{figure*}

In order to find the best-fit parameters from photoionization models,
we run a grid of models with different values of ionization parameter and 
hydrogen density. 

An attempt to use as the ionizing source a Kurucz continuum with temperatures
up to 50\,000 K confirmed that in this case it is not possible to fit the 
\ion{[O}{i]}/\ion{H}{$\alpha$} ratio. We then took as the ionizing source a 
 power--law continuum ({\sc table power law} option in 
{\sc Cloudy}), $F_\nu \propto \nu^{-1.5}$.

Starting from the solar abundances given in CLOUDY
\footnote{e.g. H: 1.00, He: 0.1, C/H: 3.55 $\times 10^{-4}$, 
N/H: 9.33 $\times 10^{-5}$, S/H: 1.62 $\times 10^{-5}$, 
O/H: 7.41 $\times 10^{-4}$},  
abundances of all elements heavier than helium were scaled as $Z/Z_{\sun}$,
with the exception of nitrogen which was scaled as $(Z/Z_{\sun})^2$.

 We considered
acceptable a model if line ratios (\ion{H}{$\beta$}=1) were consistent within 50\% 
with the observed values (see Table~\ref{tab:ratios} for results). 
In the nucleus, photoionization by stars is ruled 
out by the strong \ion{[O}{i]} $\lambda$6300 emission (\ion{[O}{i]}/H$\alpha \ge 0.1$) which
indicates ionization by a non-thermal source. 
A simultaneous fit of all the line fluxes, in particular of the low \ion{[N}{ii]} 
$\lambda$ 6583 (\ion{[N}{ii]}/H$\alpha < 0.3$), is obtained with the following parameters: 
$Z/Z_{\sun}$ = 0.5,  $n_H$ = 100 cm$^{-3}$, $\log U$ = -3.4. 

In the regions located in the d,e knots, the best fit was obtained using as 
input ionizing source a Kurucz model atmosphere with $T = 45\,000$ K,  
$\log n_H$ = 2. $\log U$ = -3, $Z/Z_{\sun}$ = 0.3.

\subsection{Photoionization + shock models}

The presence of the tidal tails strongly suggest that Mkn 298 could be a merging system leading to  
collisional processes. 
A more realistic model of the emission line ratios should then probably take into account 
the coupled effect of a photoionizing radiation flux from an external source and  of shocks.
Therefore, we have adopted the code {\sc Suma} \citep[][ and references 
therein]{viegas, cv01a,cv01b} to  model the spectra. {\sc Suma} allows to 
calculate the spectra emitted from a gas
which is heated and ionized by  the coupled effect of a photoionizing 
radiation flux from an external source (active nucleus or stars) and  
of shocks.
The  input ionizing source is a power-law in the AGN case. Radiation from a 
stellar cluster is used for starbursts.
The input parameters are those which refer to the shock, the shock velocity, \Vs, the preshock
density, \n0, and the preshock magnetic field, \B0, and those referring to the radiation flux,
the radiation intensity \Fh (in units of number of photons cm$^{-2}$ s$^{-1}$ eV$^{-1}$
at 1 Ryd), and the spectral index $\alpha$, in the power-law case.
In case of starbursts, the parameters referring to radiation are the age of 
stellar population, $t$ in Myr, and the ionization parameter, $U$.
Moreover, the input parameters include the geometrical thickness of the clouds, D, and the relative 
abundances of He, C, N, O, Ne, Mg, Si, S, Cl, Ar, and Fe  to H.
A pre-shock magnetic field \B0=10$^{-4}$ gauss was used for all models.

The shocked clouds are moving
outward from the nucleus and radiation reaches the inner edge of the clouds opposite to the 
shock front. 


A first set of models was computed by adding a shock component to 
the active nucleus.
The best fit of the observed line  ratios was obtained with the following 
parameters (Table~\ref{tab:ratios}):  $V_{\rm s}$ = 110 km/s, $n_{\rm 0}$ = 100 cm$^{-3}$, Log $F_{\rm H}$ = 7.15, $D=0.36$ pc. 
The abundances are nearly 
solar, with the exception of sulphur: N/H=$9 \times 10^{-5}$,  
O/H=$6.6 \times 10^{-4}$; S/H= $1.1 \times 10^{-5}$,

In the case of a starburst + shock,
the best fit was obtained with: 
N/H=$3.5 \times 10^{-5}$,  O/H=$6.3 \times 10^{-4}$, S/H= $5 \times 10^{-6}$
(Z/Z$_{\sun} \sim 0.6$);
$\log U=-2.52$, $t=0.0-2.5$ Myr, $D=0.5$ pc, 
$V_{\rm s}$ = 80 km/s, $n_{\rm 0}$ = 60 cm$^{-3}$.
Such values of the shock velocity and  preshock density are 
typical of the bulk of ionized gas in starburst galaxies \citep{viegas99}.

While in both cases the models provide a good fit of the observed line ratios,
the flux from the active nucleus is extremely low, a factor $\sim$ 1000 
lower than what is derived e.g. in Seyfert 2s, Log $F_{\rm H} > 11$
or in Liners, Log $F_{\rm H} \ge 8$ \citep{cvc}.
Moreover, the \ion{[O}{iii]}$\lambda$4363/\ion{H}{$\beta$} ratio is 
overpredicted by the AGN+shock models by a factor $\sim$ 10.

We conclude that while the
presence of an active nucleus cannot be excluded, it is not required in order 
to explain the observed line ratios, assuming that a shock component is also
present. Ionization by a starburst, rather than by an active nucleus, 
appears to be also in better agreement with (i) the lack of well defined 
spatial trend of the line ratios with the distance from the putative active 
nucleus and (ii) the morphology of the nucleus derived from the images in 
Sec.~\ref{sec:morphology}.
Figure~\ref{fig:shmap} displays the distribution of the ionization parameter 
$U$ and of the preshock density, \n0, which result from 
the best fit of the line ratios displayed in Figure~\ref{fig:maps}: we 
obtain    -2.00 $\le \log U \le$ -2.52,  70 $\le$  \n0 $\le$ 100, with 
higher preshock densities  in N1. This confirms that the starburst+shock 
processes are spread over the whole nuclear region of Mkn 298.


The spectra of the d,e knots are all typical of  starburst regions
with \Vs=90 \kms, \n0 = 50 \cm3, $\log U \sim -2.3$ and an age of the
stellar cluster $t<$ 2.5 Myr. 
The best fit was obtained 
with N/H=$1.5 \times 10^{-5}$,  O/H=$3.3 \times 10^{-4}$; 
S/H= $3.0 \times 10^{-6}$, which correspond to Z/Z$_{\sun} \sim 0.40$.

\section{Conclusions\label{sec:conclusions}}

In this paper we have presented new imaging and integral-field spectroscopic
data analysis of Mkn 298, aimed to investigate its controversial nature.
By means of ground-based broad-band R and HST band V images we have 
performed a
detailed analysis of the morphology of this galaxy, obtaining that Mkn 
298 is an
early-type spiral, likely a S0 galaxy (B/T $\sim 0.78$), with a de 
Vaucouleurs
bulge and without any evident nuclear point-like source.
In addition, we have pointed out the presence of weak stellar shells 
around the
galaxy. These shells, together with the bright and extended eastern 
tail, and
the nuclear dust lane, contribute to strongly support the idea that Mkn 
298 is
the result of a major merger event.
Ionized gas in Mkn 298 is distributed in two or maybe three main knots, 
visible
in the H$\alpha$-continuum subtracted image, and whose spectra show 
mostly the
typical emission lines produced by thermal ionization.
This gas is extended and oriented as the dust lane, suggesting that 
these two
features are likely to be connected. The kinematic investigation of 
H$\alpha$
and \ion{[O}{iii]} velocity fields show clearly that the gas is rotating 
with a
maximum deprojected velocity of $+ 250$ \kms with respect to the systemic
velocity of the galaxy. Therefore, the integral-field data confirm the 
previous hypothesis by \citet{moles} that a disk
dominated galaxy is the secondary component of the merger event.

The analysis of the emission line ratios shows that both AGN 
photoionization models and starburst+shock models 
with sub-solar metallicity ($Z/Z_{\sun}$ $\le$ 0.6) are able to fit the 
observed values. 
In particular the high \ion{[O}{i]}/H$\alpha$, which could indicate 
the presence of an AGN, is also well reproduced by starburst+shock
models, where photoionization is given by a young ($t < $ 2.5 Myr) stellar 
population  and shocks are produced in low velocity$-$density clouds 
($V_{\rm s}$ = 80 km/s, $n_{\rm 0}$ = 60 cm$^{-3}$). AGN+shock models
are ruled by the \ion{[O}{iii]}$\lambda$4363/\ion{H}{$\beta$}
ratio which is too high if compared with observed values and by the flux
from the AGN which is unrealistically low.

Considering that emission line ratios are constant over all the 
regions where they are detected, with 
no well defined spatial trend, we conclude that the starburst+shock 
models provide a more likely possibility than an AGN: merging would induce
both collisional processes and star formation in the gas, thus producing the 
observed line ratios.
No evidence was found for the very low metallicity  claimed by 
\citet{moles}, Z/Z$_{\sun}$ $\le$ 0.1. 
We do not confirm the high SFR ($\sim$ 90 $M_{\sun}$/yr) found by 
\citet{moles}; we find instead that  SFR/\ion{H}{$\alpha$} $\le$ 
10 $M_{\sun}$/yr. A comparable SFR was found taking the IRAS infrared 
luminosity at the position of Mkn 298, even if this value is 
very uncertain.

As it concerns the knots in 
the tidal tail (d,e), the presence of shocks is not so compelling as both
pure photoionization from stars and photoionization+shock models provide a 
good fit of the observed ratios. A sub-solar metallicity 
($Z/Z_{\sun}$ $\le$ 0.4) is found in these regions as well.

\vskip 50pt

\begin{acknowledgements}
We are grateful to the referee, P.-A. Duc, for his comments that 
improved the paper.

This paper is partly based on observations made with the NASA/ESA Hubble 
Space Telescope, obtained 
from the data archive at the Space Telescope Science Institute. 
STScI is operated by the Association of Universities for Research in 
Astronomy, Inc. under NASA contract NAS 5-26555.
This research made use of the NASA/IPAC Extragalactic Database 
(NED) which  
is operated by the Jet Propulsion Laboratory, California Institute of
Technology, under contract with the National Aeronautics and Space 
Administration.
This research has also made use of the NASA/ IPAC Infrared Science Archive, 
which is operated by the Jet Propulsion Laboratory, California Institute of 
Technology, under contract with the National Aeronautics and Space 
Administration.
\end{acknowledgements}

\label{lastpage}

\end{document}